\documentclass[aps,prl,floatfix,onecolumn,reprint,amsmath,amssymb,superscriptaddress]{revtex4-1}
\usepackage{amsfonts}
\usepackage{mathrsfs}
\usepackage{amsmath}% needed for subequations
\usepackage{color}
\usepackage{natbib}
\usepackage{graphicx}
\usepackage{bm}% bold maths
\usepackage{amssymb}
\usepackage{mathptmx}
\usepackage{stmaryrd}
\usepackage{xspace}
\usepackage{epstopdf}
\usepackage{dcolumn}% Align table columns on decimal point
\usepackage{longtable}
\usepackage{multirow}
\usepackage{bbding}
\usepackage{amsmath}

\usepackage[colorlinks=true, letterpaper=true, pdfstartview=FitV, linkcolor=blue, citecolor=blue, urlcolor=blue]{hyperref}

\begin{document}

\title{Strain-driven phonon topological phase transition impedes thermal transport in titanium monoxide}

\author{Xin Jin} 
\altaffiliation{X. J. and D.-S. M. contributed equally to this work.}
\affiliation{College of Physics and Electronic Engineering, Chongqing Normal University, Chongqing 401331, China}
\affiliation{College of Materials Science and Engineering, Chongqing  University, Chongqing 400044, China}

\author{Da-Shuai Ma}
\altaffiliation{X. J. and D.-S. M. contributed equally to this work.}
\affiliation{Department of Physics \& Chongqing Key Laboratory for Strongly Coupled Physics, Chongqing University, Chongqing 400044, P. R. China}
\affiliation{Center of Quantum materials and devices, Chongqing University, Chongqing 400044, P. R. China}

\author{Peng Yu} 
\affiliation{College of Physics and Electronic Engineering, Chongqing Normal University, Chongqing 401331, China}

\author{Xianyong Ding}
\affiliation{ Department of Physics \& Chongqing Key Laboratory for Strongly Coupled Physics, Chongqing University, Chongqing 400044, P. R. China}
\affiliation{Center of Quantum materials and devices, Chongqing University, Chongqing 400044, P. R. China}

\author{Rui Wang}
\affiliation{ Department of Physics \& Chongqing Key Laboratory for Strongly Coupled Physics, Chongqing University, Chongqing 400044, P. R. China}
\affiliation{Center of Quantum materials and devices, Chongqing University, Chongqing 400044, P. R. China}

\author{Xuewei Lv}
\email[]{lvxuewei@163.com}
\affiliation{College of Materials Science and Engineering, Chongqing University, Chongqing 400044, China}
\affiliation{Chongqing Key Laboratory of Vanadium-Titanium Metallurgy and Advanced Materials, Chongqing University, Chongqing 400044, China}

\author{Xiaolong Yang}
\email[]{yangxl@cqu.edu.cn}
\affiliation{ Department of Physics \& Chongqing Key Laboratory for Strongly Coupled Physics, Chongqing University, Chongqing 400044, P. R. China}
\affiliation{Center of Quantum materials and devices, Chongqing University, Chongqing 400044, P. R. China}

\footnotetext[3]{These authors contributed equally to this work.}

%\pacs{73.20.At, 71.55.Ak, 74.43.-f}

\begin{abstract}
Topological phonon states in crystalline materials have attracted significant research interests due to their importance for fundamental physical phenomena, yet their implication on phonon thermal transport remains largely unexplored. Here, we use rigorous density functional theory calculations and symmetry analyses to explore topological phonon phase transitions under uniaxial strains and their tuning effects on thermal transport in titanium monoxide (TiO). Our calculation shows that application of 10$\%$ tension significantly diminishes lattice thermal conductivity of TiO by 77$\%$ and 66$\%$ along the $a$ and $c$ axes, respectively, at room temperature. This suppression is found to result largely from the breaking of symmetry protected degeneracy of acoustic branches, which induces a substantial enhancement of phonon scattering phase space due to the easier fulfillment of scattering selection rules. Our study provides evidence for the importance of phononic band topology in modulating thermal conductivity and offers a promising route towards controlling solid-state heat transport.
\end{abstract}
 
%Use showkeys class option if keyword %display desired

\maketitle

\section{Introduction}
The emergence of topological quantum states is one of the most prominent advancements in condensed matter physics \cite{RevModPhys.83.1057,doi:10.1126/science.1231473,RevModPhys.89.040502,10.1093/nsr/nwy142,PhysRevX.11.031050}. 
The concepts of topological electronic states have been generalized to bosonic systems, leading to the birth of topological phonons \cite{PhysRevLett.117.068001,https://doi.org/10.1002/adfm.201904784,susstrunk2016classification,PhysRevLett.119.136401, RevModPhys.84.1045,huber2016topological,10.1093/nsr/nwx086,PhysRevB.96.094106,PhysRevLett.120.016401}. 
Recently, topological phonons have received increasing research interest in both experiments and theories \cite{stenull2016topological,zhang2018double,lv2021experimental,osterhoudt2021evidence,ren2022topological}, since it is believed that the phonon can be an ideal platform for realizing topological quantum states due to its bosonic nature and hard-to-break time-reversal symmetry \cite{liu2020topological,doi:10.1063/5.0095281}. 
In addition, phonons, energy quanta of lattice vibration, also play an essential role in fundamental physical phenomena including heat conduction, light-matter interaction, thermoelectrics, and superconductivity \cite{liu2018berry}. 
In this context, it is of essential importance to realize and tune topological nontrivial phonon states in certain materials, which may result in novel physical properties and promising application prospects in a wide range of fields. 
Hitherto, various types of symmetry-enforced topological phonons have been identified theoretically in thousands of materials \cite{chen2021topological,li2021computation,xu2022catalogue} , and some of them have been confirmed experimentally \cite{zhang2018double}. However, the physical consequences underlying topological states of phonons in these materials remain largely unexplored. Particularly, as one of the most exciting consequences, tuning of thermal transport by topological phase transition (TPT) in these systems is rarely studied \cite{yue2020phonon,tang2021topological}.

It is well known that phonons usually govern thermal transport in most materials, especially in nonmetals. 
In the kinetic theory of phonon gases \cite{ziman2001electrons,srivastava1990physics}, the lattice thermal conductivity ($\kappa_L$) can be expressed as the sum of contributions from all phonon modes: 
$\kappa_{L}=\sum_{\lambda}C_{\lambda}v_{\lambda}^2\tau_{\lambda}$, where $C_{\lambda}$ is the heat capacity of a phonon mode $\lambda$, $v_{\lambda}$ is the group velocity, and $\tau_{\lambda}$ is the phonon lifetime. 
All three factors that determine $\kappa_L$ are closely associated with the phonon dispersion of a material: $C_{\lambda}$ and $v_{\lambda}$ are completely determined from the phonon dispersion, while $\tau_{\lambda}$ relies largely on the phonon dispersion due to the restriction of phonon-phonon scattering selection rules \cite{lindsay2018survey,qian2021phonon}. 
In this scenario, one can utilize the phononic TPT to alter the phonon dispersion relation of a material and thus effectively regulate its heat conduction. 
However, limited by the charge neutral and spinless nature, the TPT of phonons generally cannot be achieved by conventional approaches of treating electrons \cite{liu2020topological,yue2020phonon}.
It is noted that previous studies have demonstrated TPTs of phonons induced in different approaches, including spin–lattice interactions for magnetic lattices \cite{zhang2010topological,ioselevich1995strongly}, magnetic field for ionic lattices \cite{zhang2010topological}, Coriolis field \cite{wang2015coriolis}, and optomechanical interactions \cite{PhysRevX.5.031011}. In contrast, applying external pressure or strain to break the symmetry of crystals provides an alternative and universal method for achieving TPTs of phonons \cite{jiang2019strain,rostami2022strain,katailiha2021topological}.

In this work, we perform first-principles calculations to examine the effect of phononic TPT on the thermal transport in hexagonal titanium monoxide (TiO), a Weyl semimetal that has been synthesized experimentally \cite{mohr1994neue}. With the aid of symmetry analyses, we identify the existence of phononic triple degenerated point (TDP) and the nodal line formed doubly degenerated acoustic phonon branches along the high-symmetric $\Gamma\mathrm{\bf A}$ direction in the Brillouin zone (BZ). Our calculation shows that phononic TPT can be realized by applying uniaxial strains along the [100] crystal direction, and its impact on the thermal conductivity becomes prominent with the increase of strain. Using the 10$\%$ uniaxial tension as a study case, we reveal that the TPT of phonons contributes significantly to the suppression of $\kappa_L$ along the $c$ axis. By further modal phonon transport analysis, we attribute the reduction in $\kappa_L$ to the increased phonon-phonon scattering channels, which is closely associated with the breaking of degenerate states of topological phonons. Our work reveals the strong relevance of topological phonons with thermal conductivity, and opens up a pathway towards controlling heat conduction in topological materials.

 \begin{figure*}[!htb]
    \centering
    \includegraphics[scale=0.5]{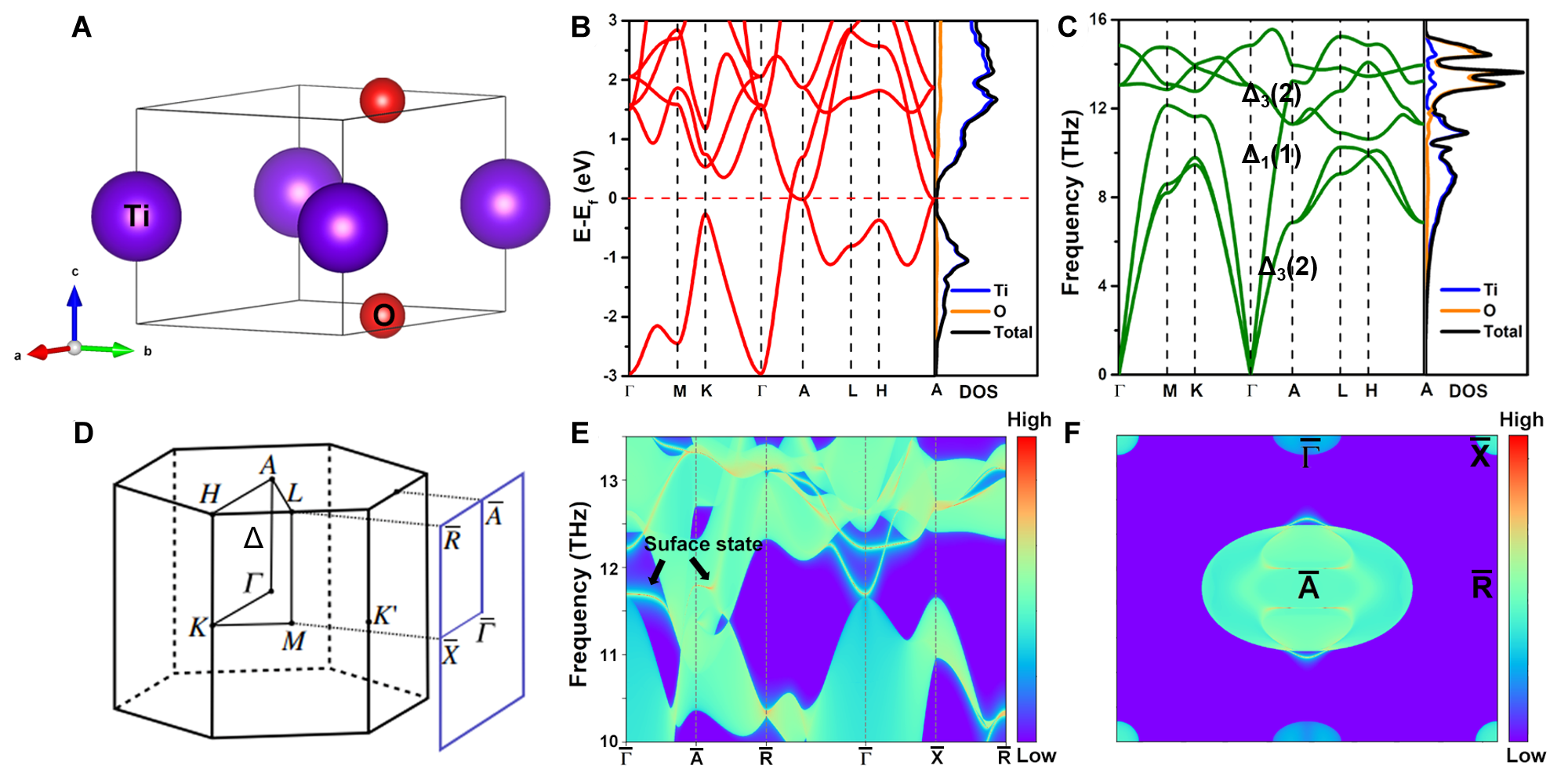}
    \caption{Crystal structure, electron band structure, phonon spectrum, BZ, and suface state. (A) Crystal structure of hexagonal WC-type TiO. The Ti atom and O atom are represented by sphere in purple and red, respectively. 
    (B) The calculated electron band structure along high-symmetry lines and corresponding electron density of states.
    (C) The calculated phonon spectrum and corresponding phonon density of states. 
    The Irreps of phonon branches along $\Delta$ are inserted in which the number in brackets indicate the dimension of the Irrep and the degeneracy fold of phonon branches along $\Delta$.
    (D) The 3D bulk BZ and projected surface BZ of (100) surface. 
    (E) The surface states of TiO on the (100) surface. 
    (F) Iso-energy band contours of (100) surface of TiO around 12~THz. The graduated color bar represents the amplitude of the projected density of states in (E) and (F).
    \label{FIG1}}
\end{figure*}

\begin{figure}[!htb]
    \centering
    \includegraphics[scale=0.095]{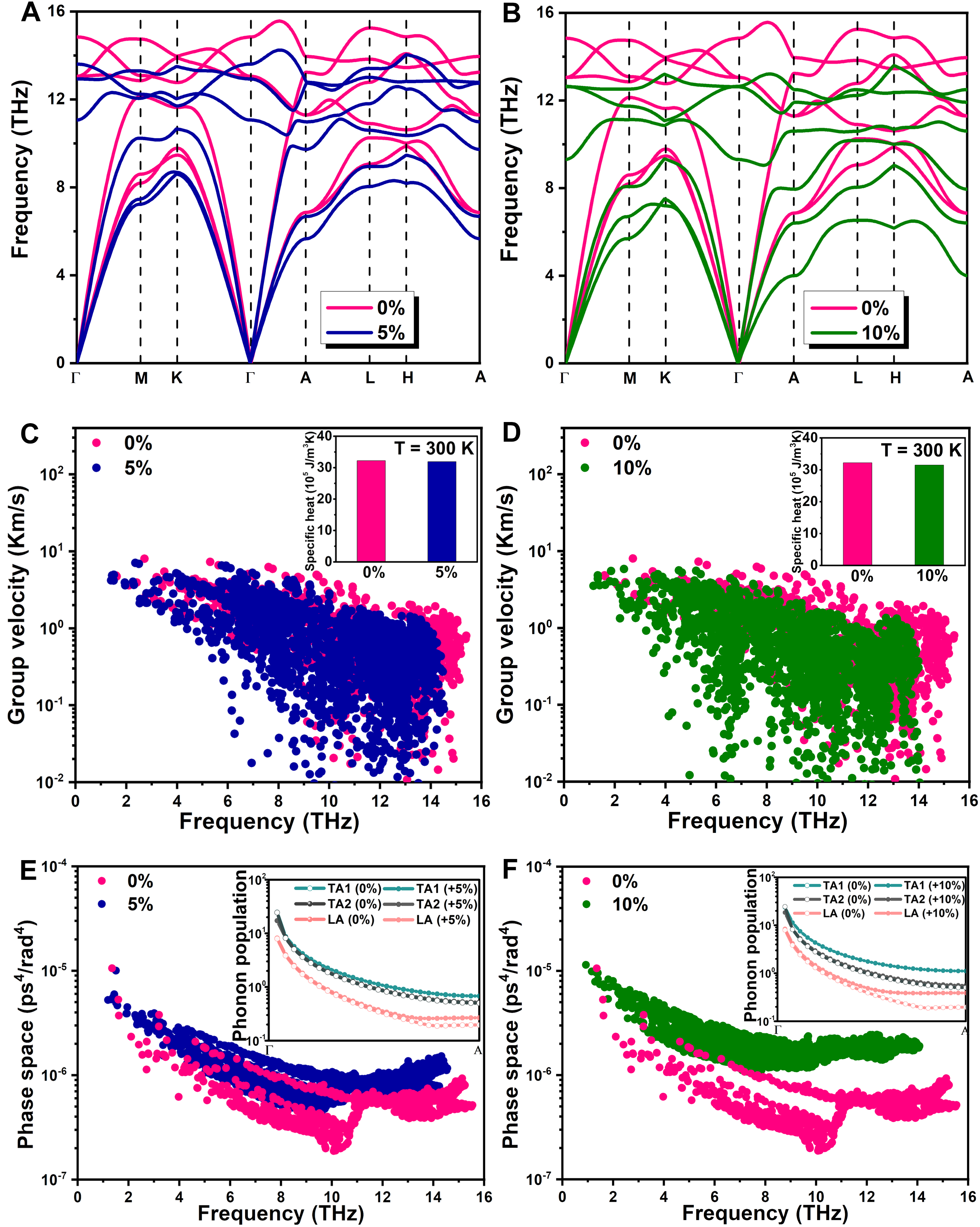}
    \caption{Phonon spectra, group velocities, and scattering phase space before and after TPT.
    (A) Phonon dispersions under uniaxial strain along the [001] direction at (A) 5$\%$ and (B) 10$\%$. Phonon group velocities under uniaxial strain at (C) 5$\%$ and (D) 10$\%$. The insert shows the comparison of volumetric heat capacity values under different strains with the intrinsic heat capacity values. Three-phonon phase space under uniaxial strain at (E) 5$\%$ and (F) 10$\%$. Additionally, the inset depicts the phonon population at 300 K. 
     \label{FIG2}}
\end{figure}
 
\section{Results}
\subsection{Topological characteristics of phonons}
%\subsection{Symmetry protected degeneracy in the phonon spectrum of TiO}
TiO crystallizes in the tungsten carbide type (WC-type) hexagonal structure with space group of $P\overline{6}m2$ (No.~187), and has recently been predicted to have nontrivial topological phonon states \cite{PhysRevB.99.174306}. 
The crystal structure of WC-type TiO is shown in Fig.~\ref{FIG1}(A), where the Ti atom and the O atom occupy the wyckoff position $1b$ (0, 0, 1/2) and $1c$ (1/3, 2/3, 0), respectively. 
The topological electronic states in TiO have been well studied \cite{ULLAH2022}. 
As is seen in Fig.~\ref{FIG1}(B), in the electronic band of TiO, there exists a TDP along the high-symmetry line $\Gamma \mathrm{\bf A}$  that brings in the vanishing of the density of states near the Fermi level, revealing the nature of its semimetal. Fig.~\ref{FIG1}(C) shows the calculated phonon dispersion of TiO along the high-symmetry lines along with corresponding phonon density of states. 
The missing of imaginary frequency modes indicates the dynamical stability of TiO. 
From the phonon density of states, it is also observed that the low-frequency phonon modes below 12 THz are dominated by Ti atoms, while those above 12~THz mainly originate from the lighter O atoms. 
Notably, we observe rich types of degeneracy of phonon branches in the phonon spectrum of TiO. 
Specifically, two transverse acoustic (TA) branches degenerate along the $\Gamma\mathrm{\bf A}$ line (marked as $\Delta$), forming a nodal line running through the BZ, and the longitudinal acoustic (LA) branch crosses with a double-degenerated optical branches along $\Delta$ forming a TDP at $\sim$12~THz. The emergence of these types of phonon band degeneracy is explained as follows.

In terms of symmetry analysis, the generating elements of the little group at an arbitrary point in $\Delta$ are $C_{3z}^+$, $M_{100}$ associate with joint symmetry $M_{001}T$. For the degenerated TA branches, the eigenvalues of $C_{3z}^+$ symmetry are $\left[e^{-2\pi i /3},e^{2\pi i /3}\right]$. 
Due to the existence of the joint symmetry $M_{001}T$ at any arbitrary $k$ point on $\Delta$, these two symmetry eigenvalues that are complex conjugates of each other must be degenerated forming a two-dimensional irreducible representation (Irrep) $\Delta_3$ written under the BCS convention \cite{Elcoro:ks5574,Aroyo:xo5013}. 
Thus, the double-fold degenerated nodal line is protected by $M_{001}T$ and $C_{3z}^+$. 
As for the LA phonon branch, its Irrep is $\Delta_1$ and the eigenvalue of $C_{3z}^+$ symmetry is 1. The similar symmetry analysis is also applied to the three optical branches, and corresponding Irreps are labeled in Fig.~\ref{FIG1}(C). Notably, the TDP at 12~THz is formed by two sets of bands ($\Delta_1$ and $\Delta_3$) with different $C_{3z}^+$ eigenvalues, indicating that this degeneracy is protected by $C_{3z}^+$ and $M_{001}T$. 
Since there does not exist a fully gapped sphere surrounding 
the TDP, this TDP does not have a well-defined topological charge. Even though, this TDP can be regarded as two Weyl points with opposite topological charges $\pm1$ and formed by band pairs whose $C_{3z}^+$ eigenvalues are $\left[e^{2\pi i /3},1\right]$ and $\left[e^{-2\pi i /3},1\right]$ degenerate together. 
As a result, when projected onto surface, one can observe the Fermi arc whose origin and terminal are located at the same projected TDP. To verify this point, the density of states of (100) surface of TiO is calculated and shown in Fig.~\ref{FIG1}(E). One notices that the surface state is clearly visible. 
To understand the connection of the surface state clearer, we also plot the Fermi surface of the phonon dispersion of the semi-infinite (100) surface in Fig.~\ref{FIG1}(F). As expected, the Fermi arc origins and terminate exist at the same point in the projected BZ. 

\begin{figure}[!htb]
    \centering
    \includegraphics[scale=0.11]{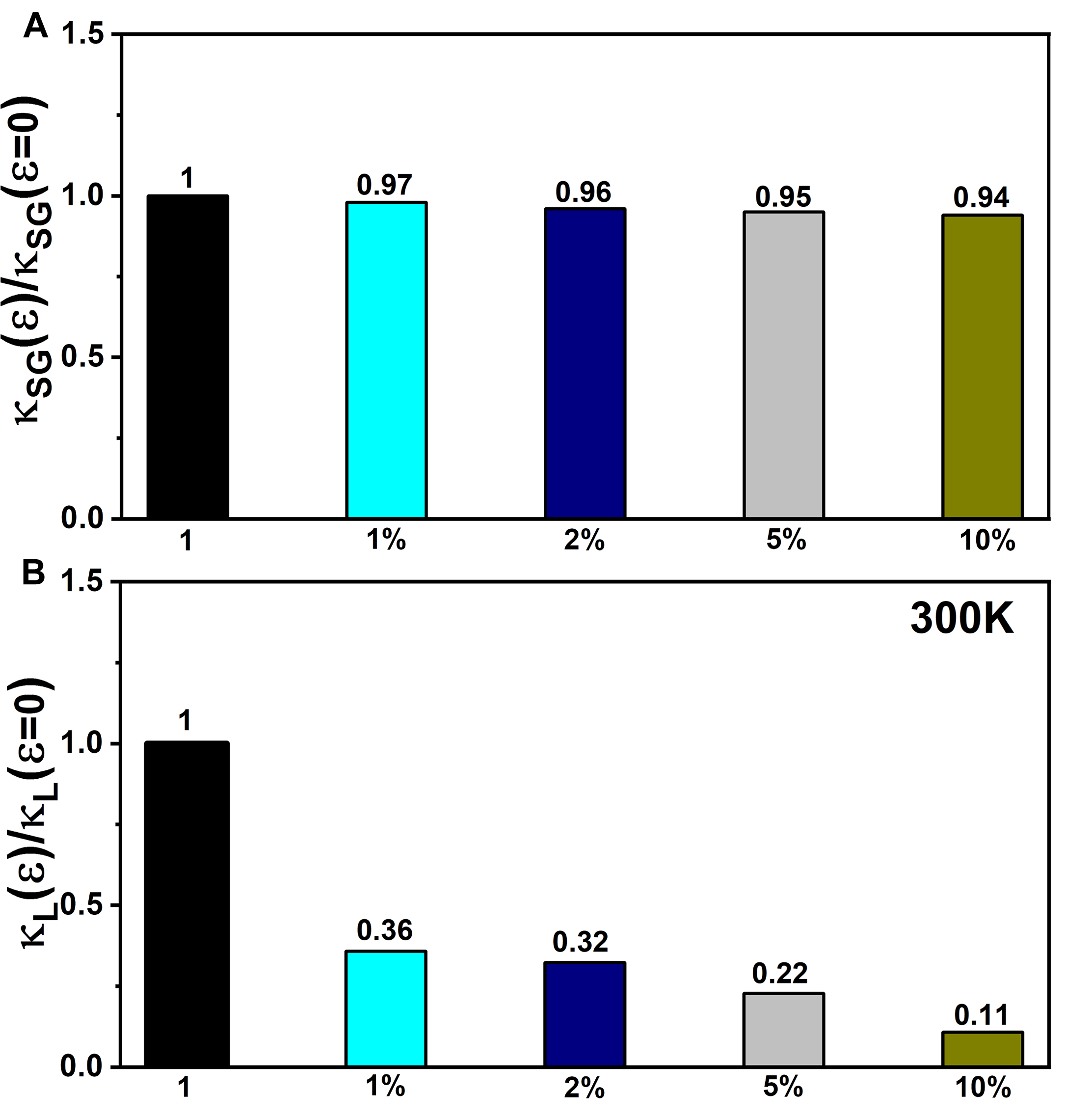}
    \caption{Normalized $\kappa_{\rm SG}$ and $\kappa_{L}$. (A) Normalized $\kappa_{\rm SG}$ along the $c$ axis at different strains at 300 K. (B) Normalized $\kappa_{L}$ along the $c$ axis at different strains calculated using the anharmonic IFCs at $\epsilon$ = 0. $\kappa_{\rm SG}$ and $\kappa_{L}$ are normalized by the unstrained values. 
     \label{FIG3}}
\end{figure}

\begin{figure}[!htb]
    \centering
    \includegraphics[scale=0.10]{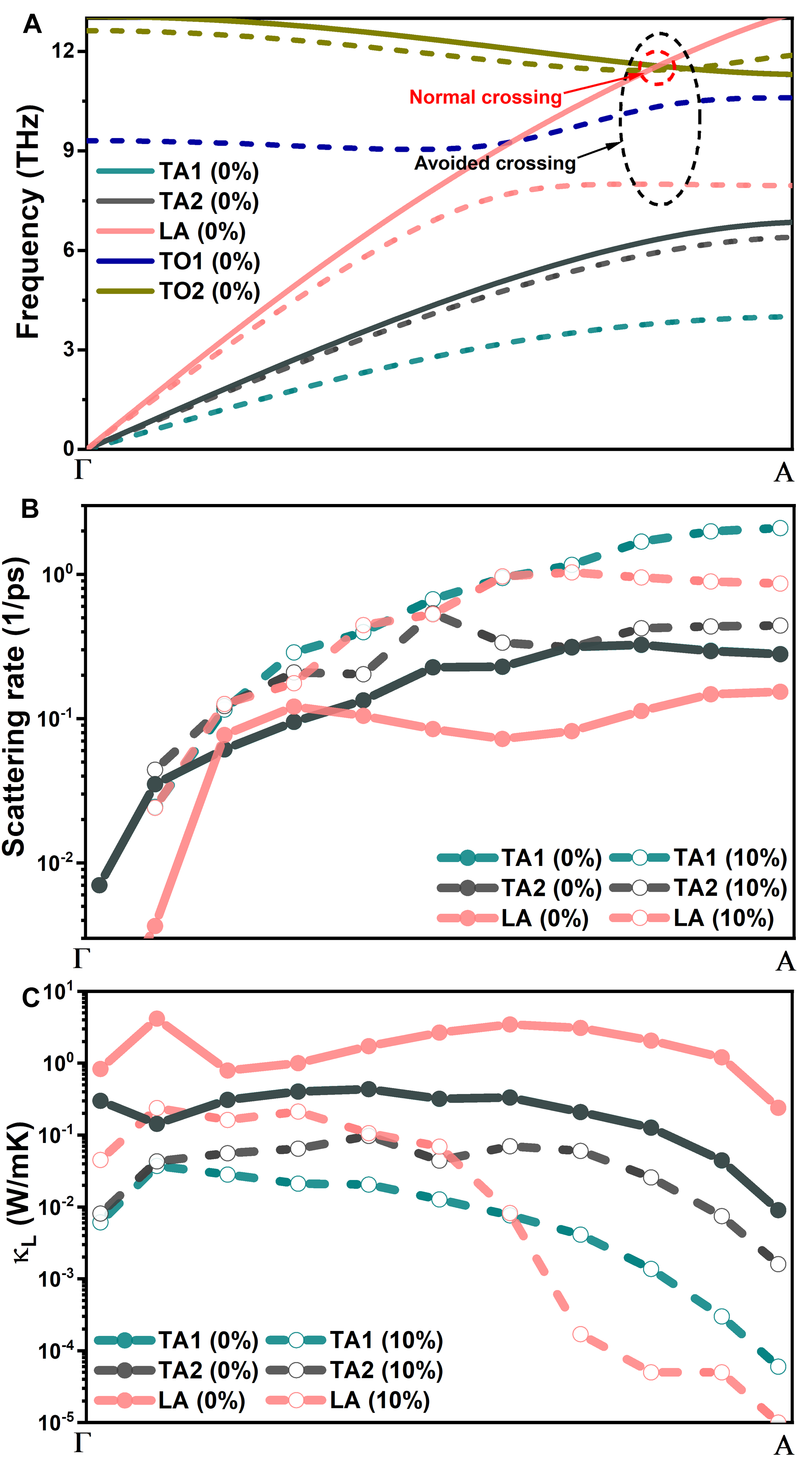}
    \caption{Phonon dispersion curves, scattering rates, and $\kappa_{L}$ along the $\Gamma \mathrm{\bf A}$ line. (A) The phonon dispersion curves along the $\Gamma \mathrm{\bf A}$ line for unstrained and 10$\%$ strained cases. Comparison for the avoided-crossing behaviour is marked by the dashed circles. (B) Comparison of scattering rates for acoustic phonons at 300 K under different strains. (C) Comparison of $\kappa_{L}$ values at 300 K along the $\Gamma \mathrm{\bf A}$ line. 
     \label{FIG4}}
\end{figure}
\subsection{Strain-induced topological phonon phase transition and its impact on thermal conductivity}
The presence of these phonon band topologies should have non-negligible impact on thermal conductivity, since they can profoundly affect group velocities and phase space for phonon-phonon scatterings. To explore their effect on thermal transport, we attempt to achieve topological phonon phase transitions by applying strain. By symmetry analysis, we find that applying an uniaxial strain along the [100] crystal direction can break the phonon band degeneracy protected by the $C_{3z}^+$ and $M_{001}T$ symmetries. To this end, we study the evolution of phonon dispersion under a series of uniaxial (100) tensile strains. Phonon spectra of TiO under two representative tensile ($\epsilon$ = 5$\%$, 10$\%$) strains are compared with the unstrained case in Fig.~\ref{FIG2}(A, B). It is seen that with the presence of tensile strain, the degenerated TA branches along the $\Gamma \mathrm{\bf A}$ direction are separated, and the degree of degenerate opening gradually increases with strain. Notably, in the case of 10$\%$ tensile strain, the significant softening of the low-lying TA modes along the $\Gamma \mathrm{\bf A}$ direction leads to the large separation of three acoustic branches, which allows for increased phonon-phonon scattering channels due to the easier fulfillment of energy and momentum conservation rules \cite{PhysRevB.100.245203,PhysRevB.101.161202,ravichandran2019non}. Simultaneously, applying strain makes the TDP vanish as well, accompanied by the appearance of the avoided crossing between LA and TO phonon branches, which facilitates to enlarge the phase space for available phonon-phonon scatterings as mentioned previously \cite{delaire2011giant,lu2018lattice}. Besides, the phonon dispersion curves gradually soften as the tensile strain increases, corresponding to lower group velocities in Fig.~\ref{FIG2}(C, D), which is capable of reducing $\kappa_L$. Additionally, it can be seen from the insets of Fig.~\ref{FIG2}(C, D) that the heat capacity, another important factor determining $\kappa_L$, is only slightly affected by strains.  

\begin{figure}[!htb]
    \centering
    \includegraphics[scale=0.105]{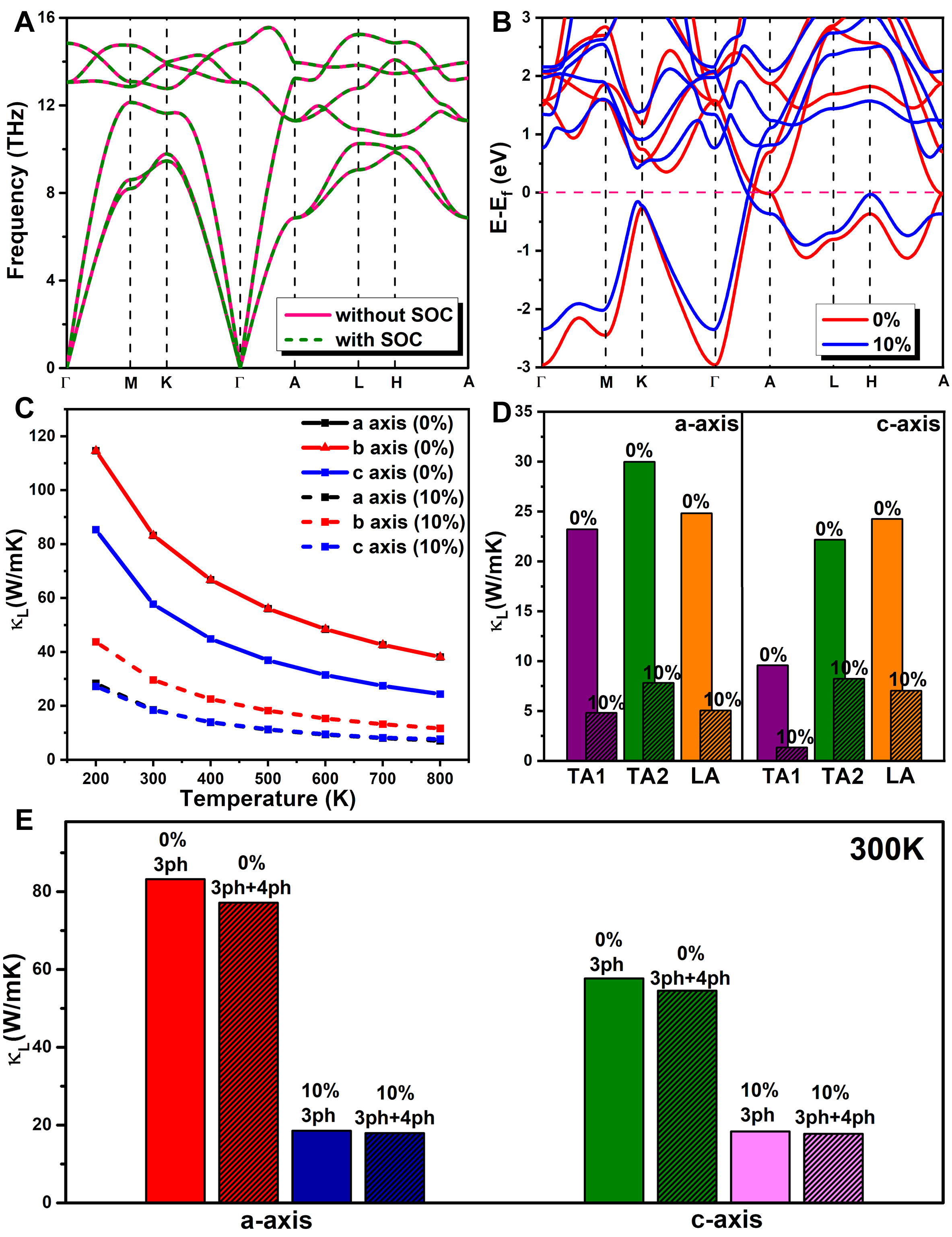}
    \caption{Electron band structure and $\kappa_{L}$ before and after TPT. (A) The phonon dispersion of the intrinsic TiO with and without considering SOC. (B) Electronic band structures at 0 and 10$\%$ strains. (C) Temperature-dependent $\kappa_{L}$ of TiO at different strains along the $a$, $b$, and $c$ axes. (D) The contribution to $\kappa_L$ from TA1, TA2, and LA modes at 0 and 10$\%$ strains along the $a$ and $c$ axes. The inset is the values of $\kappa_{L}$ at 300 K. (E) The calculated lattice thermal conductivity with (3ph+4ph) and without (3ph) four-phonon scattering included along $a$ and $c$ axes at room temperature.
     \label{FIG5}}
\end{figure}

\begin{figure*}[!htb]
    \centering
    \includegraphics[scale=0.13]{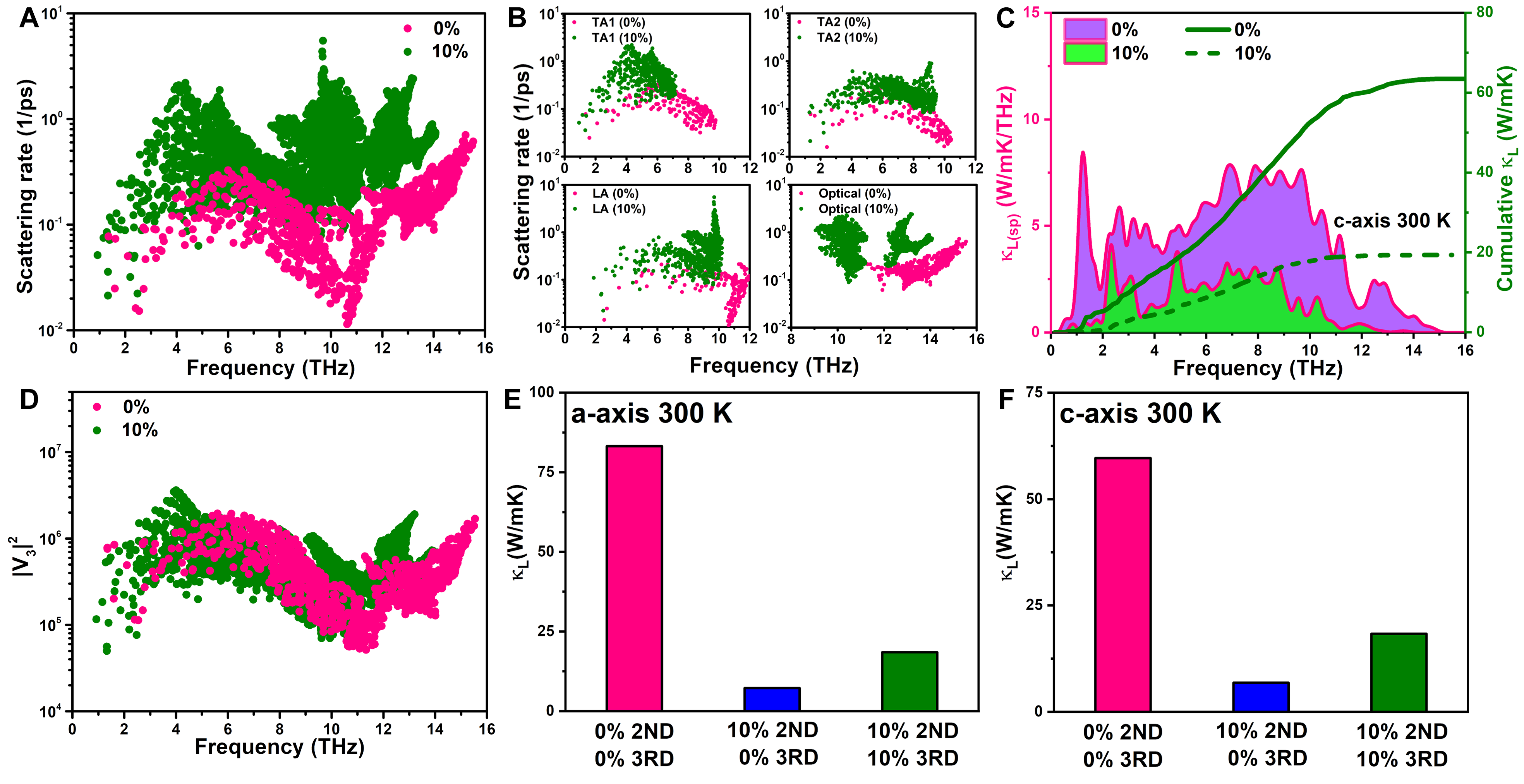}
    \caption{Phonon scattering rates, cumulative $\kappa_{L}$, and phonon anharmonic before and after TPT. (A) The total phonon scattering rates at $T$ = 300 K. (B) The phonon scattering rates from different branches at $T$ = 300 K. (C) Spectral and cumulative $\kappa_{L}$ along the $c$-axis with respect to the phonon frequency. (D) The transition probability matrices $\left|{V}_{3}\right|^{2}$. The calculated $\kappa_{L}$ from the mixed IFCs along (E) $a$ and (F) $c$ axes.
    \label{FIG6}}
\end{figure*}

From the perspective of phonon scattering, the prominent changes in the phonon dispersion revealed above are expected to affect phonon lifetimes by altering the phonon-phonon phase space. Fig.~\ref{FIG2}(E, F) shows the corresponding three-phonon phase space under different strains. It is evident that the modal level phase space is largely increased when the tensile strains are applied, especially for the 10$\%$ case, which should be attributed to the combined effect of increased phonon population and TPT-induced selective rules for phonon scattering processes. On the one hand, tensile strains make the overall phonons shift to lower frequencies due to the phonon softening, which results in the increased phonon population as seen in the insets in Fig.~\ref{FIG2}(E, F) and thus facilitates to increase the phase space. On the other hand, strain-induced phonon TPT makes the energy selection rule of three-phonon processes easier satisfied, which also tends to increase the phase space. 
 
%\subsection{ Topological phonon phase transition by uniaxial strain}
\begin{figure*}
    \centering
    \includegraphics[scale=0.13]{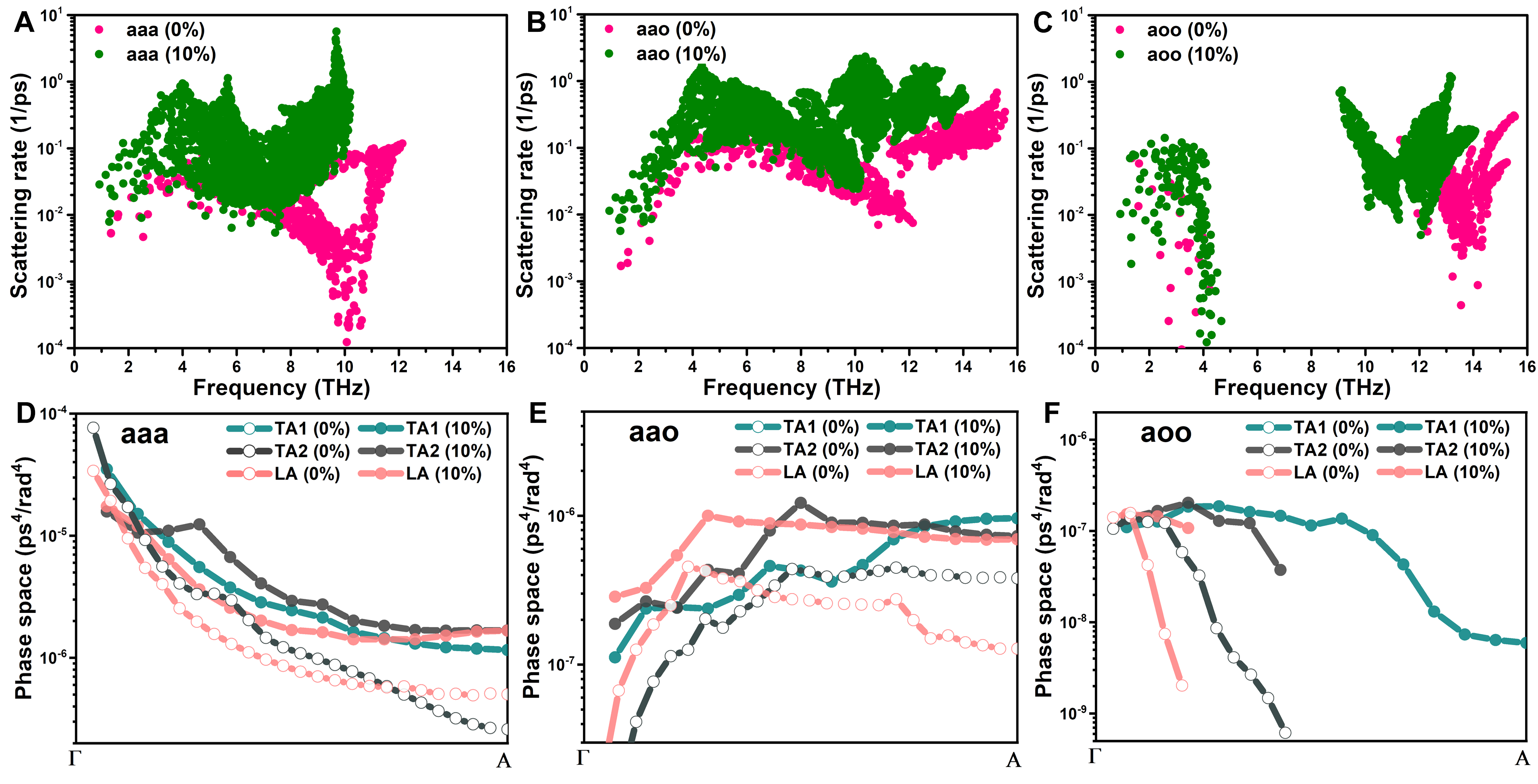}
    \caption{Phonon scattering rates and phase spaces from different processes. The contribution to (A-C) phonon scattering rates and (D-F) weighted phase spaces along the $\Gamma \mathrm{\bf A}$ line from different processes. 
    \label{FIG7}}
\end{figure*}

To quantify the impact of strain-induced changes in the phonon dispersion on thermal transport, we calculate the small-grain-limit reduced thermal conductivity, $\kappa_{\rm SG}$, at different strains. $\kappa_{\rm SG}$ can measure the influence of the harmonic properties on $\kappa_{L}$, which is defined as $\kappa_{\rm SG}=\sum_{\lambda}C_{\lambda}v_{\lambda}^2$. Since the strain-induced phononic TPT occurs in the $\Gamma \mathrm{\bf A}$ direction, which is parallel to the $c$ axis and perpendicular to the $a$ and $b$ axes as judged from the BZ in Fig.~\ref{FIG1}(D), it is highly associated with the thermal transport along the $c$ axis. In view of this, we mainly focus on the effect of phonon TPT on the thermal conductivity along the $c$ axis. Fig.~\ref{FIG3}(A) shows the normalized $\kappa_{\rm SG}$ along the $c$ axis using the intrinsic case as a criterion for different strains. It can be seen that the variations in $\kappa_{\rm SG}$ are less sensitive to strains, which actually reflect the influence of changes in group velocities and heat capacity on $\kappa_{L}$. Specifically, tensile strains slightly decrease $\kappa_{\rm SG}$, e.g., only 6$\%$ reduction under 10$\%$ tension. This indicates that the change of harmonic transport properties due to phonon softening has a rather weak effect on the thermal conductivity.

In addition to the harmonic properties, strain also affects phonon scattering strength by changing the anharmonic IFCs, which determines phonon-phonon scattering matrix elements. The phonon scattering strength is determined jointly by the phase space and the scattering matrix elements. In order to emphasize the consequences from changes in the phase space, we calculate the thermal conductivities under different strains, assuming that the anharmonic IFCs remains unchanged. Fig.~\ref{FIG3}(B) displays the normalized $\kappa_{L}$ along the $c$ axis at 300 K using the unstrained value as a reference. The effect of phase space on $\kappa_{L}$ can then be isolated by analyzing the difference between the normalized $\tilde{\kappa}_{\rm SG}=\kappa_{\rm SG}(\epsilon)/\kappa_{\rm SG}(\epsilon=0)$ and normalized $\tilde{\kappa}_{L}=\kappa_L(\epsilon)/\kappa_L(\epsilon=0)$, namely $\Delta \tilde {\kappa}=|\tilde{\kappa}_{L}-\tilde{\kappa}_{\rm SG}|$. It is found that $\Delta \tilde{\kappa}$ is substantially larger than the change in $\kappa_{\rm SG}$ for all strain cases, and $\Delta \tilde{\kappa}$ becomes larger as the strain increases, which is enabled by the increased phase space due to the enlarged separation of three acoustic branches by increasing the strain. For the 10$\%$ case in particular, $\Delta \tilde{\kappa}$ = 0.83, indicating that the phase space change reduces $\kappa_{L}$ by over 80$\%$. This conveys to us that the effect of strain on $\kappa_L$ is almost entirely dominated by the phase space change, which is closely connected to the phonon TPT. 

\subsection{Tensile strain-suppressed thermal conductivity}
Then we elaborate how the phonon TPT affects the thermal conductivity in more detail, taking the case of 10$\%$ uniaxial tension--where the effect of phonon TPT is most pronounced--as an example. As seen in Fig.~\ref{FIG4}(A), when 10$\%$ strain is applied, the phononic TPT occurs with the significant softening of LA branch, which leads to avoided-crossing behavior between the LA and TO branches along the $\Gamma \mathrm{\bf A}$ direction. Meanwhile, a large separation of three acoustic branches from the original degeneracy between two TA branches is clearly observed. These changes in phonon spectrum due to the TPT can bring about reduced group velocities and enlarged phonon scattering phase space, especially for the LA and low-lying TA branches when approaching $\mathrm{\bf A}$ point. It is indeed observed from Fig.~\ref{FIG4}(B) that in the presence of 10$\%$ strain, the phonon scattering rates of three acoustic branches along the $\Gamma \mathrm{\bf A}$ line are remarkably elevated due to the enlarged phase space, which is a combined result of degenerate opening between TA branches and avoided crossing between the LA and TO branches. Consequently, the corresponding modal thermal conductivity of acoustic branches is significantly suppressed, as depicted in Fig.~\ref{FIG4}(C). These results explicitly proves that phononic TPT enables the significant suppression of $\kappa_L$.  

When it comes to TPT, spin-orbit coupling (SOC) is believed to be universally important in electronic systems. A natural question is whether it also affects the phonon properties. To address this question, we study the effects of SOC on the phonon dispersion. As is seen in Fig.~\ref{FIG5}(A), SOC results in no detectable change on the phonon dispersion of intrinsic TiO, compared with that obtained without SOC. In parallel, we also pay attention to the tuning effect of strain on the electronic band structure, as this may involve the scattering of phonons by electrons. As plotted in blue in Fig.~\ref{FIG5}(B), in the presence of strain, the TDP electronic state is split into two double degenerated points, and the system remains the semimetal phase. Given the vanishing electronic density of states at the Fermi level, the phonon scattering by electrons across the TPT should thus be insignificant \cite{yang2021indirect,yang2021tuning}. 

To examine the strain effect on the phonon transport, we present the calculated temperature-dependent $\kappa_L$ under 0$\%$ and 10$\%$ strain in Fig.~\ref{FIG5}(C). In the absence of strain, the $\kappa_L$ values along the $a$ and $b$ axes are the same and larger than those along the $c$ axis. As a result of strain-induced crystal symmetry breaking, the $\kappa_L$ values along three axes are all different when TiO is subjected to 10$\%$ strain, with the one along $b$ axis being largest, followed by the $a$ axis and then the $c$ axis. Importantly, the $\kappa_L$ values along three axes under 10$\%$ strain are all significantly declined over the entire temperature range, compared with the intrinsic case. Specifically, at 300 K, $\kappa_{L}$ along the $a$, $b$, and $c$ axes decreases by 77$\%$ (from 83 to 19 W/mK), 64$\%$ (from 83 to 30 W/mK), and 66$\%$ (from 58 to 18 W/mK), respectively. Further decomposing $\kappa_L$ into different branches, we can see from Fig.~\ref{FIG5}(D) that under 10$\%$ strain the contributions to $\kappa_L$ from all three acoustic branches are substantially decreased. At 300 K, the $\kappa_L$ contributed from the TA1, TA2, and LA modes along the $a$ ($c$) axis is reduced by 80$\%$ (86$\%$), 74$\%$ (63$\%$), and 80$\%$ (72$\%$), respectively. In particular, for the intrinsic case, the topologically protected TA1 and TA2 degenerate states account for 17\% and 38\% of $\kappa_L$ along the c-axis, respectively, indicating that topological phonons  contribute directly more than 50\% to the $\kappa_L$. It should be pointed out that the values of $\kappa_L$ along different axes all drop significantly under the strain, but the reasons behind them are quite different. For the $a$ and $b$ axes, the suppression of $\kappa_L$ is largely due to the phonon softening effect caused by the strain itself, as broadly seen in many other systems \cite{zhou2022thermal,PhysRevB.106.184303,PhysRevB.90.235201}, whereas for the $c$ axis the decrease of $\kappa_L$ mainly originates from the strain-induced TPT. Note also that here the thermal conductivity is computed within the three-phonon picture due to the limitation of computational cost. To evaluate the effect of four-phonon scattering in TiO, we also calculate $\kappa_L$ at room temperature by considering phonon anharmonicity up to the fourth order. As is seen in Fig.~\ref{FIG5}(E), after including four-phonon scattering, $\kappa_L$ along the $a$ axis decreases by 7$\%$ (from 83 to 77 W/mK) at zero strain and 5$\%$ (from 19 to 18 W/mK) under 10$\%$ strain, respectively. The story is similar for the $c$ axis. This indicates that the effect of four-phonon scattering is not strong and is thus not considered in the present calculation.  

\subsection{Physical origin of thermal conductivity reduction}

The thermal conductivity of a material is jointly determined by group velocities $v_{\lambda}$, heat capacity $C_v$, and phonon lifetimes $\tau_{\lambda}$. From the above analysis of $\kappa_{\rm SG}$, we can see that the reduction in $\kappa_L$ along the $c$ axis resulting from group velocities and heat capacity accounts only for 6$\%$, far less than the actual reduction of 66$\%$. This indicates that the significant suppression in $\kappa_L$ comes mainly from the shortened phonon lifetimes. Hence, we analyze the phonon scattering rates, inverse of phonon lifetimes, as seen in Fig.~\ref{FIG6}(A). It is clear that through the TPT, the phonon scattering rates are significantly increased in the whole frequency range. Particularly, the original phonon scattering rates have a very marked dip in the frequency range of 6-12 THz, while the scattering rates of these phonons increase by almost an order of magnitude when the 10$\%$ strain is applied. This prominent feature is closely related to the enlargement of phase space for three-phonon scattering as discussed later. 

For more insights, we compare the phonon scattering rates of each branch in Fig.~\ref{FIG6}(B). It is showed that the phonon scattering rates of all the heat-carrying acoustic branches under the strain are substantially larger than those without the strain, especially for the low-lying TA modes below 6 THz. To illustrate which frequency ranges are affected across the TPT, the spectral lattice thermal conductivity along the $c$ axis and its accumulation with frequency are provided in Fig.~\ref{FIG6}(C). Indeed, phonons below 10 THz, corresponding to acoustic branches, are a major contributor to thermal conductivity and their contributions are largely diminished across the TPT. 

The phonon scattering rates depend on the phase space and the scattering matrix elements, both of which can be altered by the applied strain. To find out which one dominates the increase in phonon scattering rates, we detect these two quantities separately. Fig.~\ref{FIG6}(D) shows the mode-resolved scattering matrix elements, which can be used to measure the phonon anharmonicity \cite{jin2022bonding,zhou2022anomalous,jin2022comprehensive}. It can be seen that the mode-level scattering matrix elements, although somewhat different, are comparable in magnitude before and after applying the strain. The phonon phase space characterizes the number of all available phonon scattering channels for simultaneously satisfying the energy and momentum conservation. As seen in Fig.~\ref{FIG2}(F), the overall phase space of all phonons is significantly increased when 10$\%$ strain is applied. Thus, the increase of phase space induced by TPT is a leading driver of enhancing the phonon scattering. To quantify effects of these two quantities on $\kappa_L$, we calculate the thermal conductivity using the mixed IFCs. By combining the harmonic IFCs at $\epsilon$ = 10$\%$ and anharmonic IFCs at $\epsilon$ = 0$\%$, we obtain $\kappa_L$ = 7 and 7 W/mK along the $a$ and $c$ axes at 300 K, which is lower than the actual value of $\kappa_L$ = 19 and 18 W/mK using all the IFCs at $\epsilon$ = 10$\%$, as given in Fig.~\ref{FIG6}(E, F). This implies that the lattice anharmonicity is actually weakened after imposing the strain, and if it remains constant the effect of phonon TPT on $\kappa_L$ would be even stronger. 

To establish a deeper understanding of connection between the phonon band topology and phonon scattering rates, we further analyze the contributions of different processes to phonon scattering rates. As shown in Fig.~\ref{FIG7}(A-C), the phonon scattering rates below 10 THz are governed by three-phonon processes involving either three acoustic phonons ($aaa$) or two acoustic and one optical modes ($aao$), while processes among one acoustic and two optical phonons ($aoo$) have a minor contribution. At $\epsilon$ = 0$\%$, the double-fold degenerated nodal line between TA branches allows the acoustic phonons to bunch together, which prevents the $aaa$ processes due to the constraint of phonon scattering selection rules. As a consequence, the phonon scattering rates arising from $aaa$ processes exhibit a large dip in a wide frequency range below 12 THz, corresponding to the relatively lower phase space of $aaa$ processes along the $\Gamma \mathrm{\bf A}$ direction shown in Fig.~\ref{FIG7}(D). Upon the 10$\%$ uniaxial tension, the opening of the degenerated TA branches largely weakens the acoustic bunching effect, resulting in the significantly increased phase space of $aaa$ processes for all three acoustic branches as illustrated in Fig.~\ref{FIG7}(D). Thus, the phonon scattering rates contributed by $aaa$ processes are largely increased accordingly, as seen in Fig.~\ref{FIG7}(A). Besides, at $\epsilon$ = 10$\%$, the three optical phonon branches also show softening, accompanied by the avoiding crossing between the LA and optical phonons, as shown in Fig.~\ref{FIG4}(A), which makes it easier for $aao$ processes to satisfy the energy conservation rules. As a result, the phase space of $aao$ processes for all acoustic branches is considerably enlarged after the phonon TPT, as demonstrated in Fig.~\ref{FIG7}(E). This results in the enhancement of phonon scattering involving $aao$ processes. For $aoo$ processes, although they have only a minor contribution to scattering rates, it can be seen from Fig.~\ref{FIG7}(F) that through the TPT, their phase spaces for three acoustic phonon branches are all increased significantly. Especially for the low-lying TA (TA1) branch, when approaching the $\mathrm{\bf A}$ point, $aoo$ process cannot even occur in the unstrained case, but makes a significant contribution to the phase space after the strain is applied. These results provide strong evidence that topological effects of phonons strength phonon scattering by increasing available scattering channels and thus suppress $\kappa_L$, which can be a effective strategy to improve the performance in many applications requiring low $\kappa_L$, such as thermoelectrics and thermal barrier coatings. 

\section{Discussion}
In summary, we have disclosed significant impact of strain-induced topological phonon phase transition in suppressing the lattice thermal transport of TiO by first-principles calculations combined with the linearized phonon BTE. The systematic symmetry analyses explicitly show the coexistence of the nodal line formed by two TA branches and a TDP among the LA and optical branches. We apply a series of uniaxial strains to achieve the TPT of these nontrivial phonons, and explore its implication on the phonon thermal transport. Importantly, we find that application of the 10$\%$ uniaxial tensile action causes the large separation of acoustic branches and avoided crossing between LA and TO branches due to the breaking of degenerate states of topological phonons, which increases phonon-phonon scattering phase space drastically and consequently results in significant suppression of thermal conductivity along the $c$ axis, with 66$\%$ reduction at room temperature. This finding establishes the correlation between the phononic TPT and thermal conductivity, and provides a fundamental basis for regulating heat conduction in solids.    

\section{EXPERIMENTAL PROCEDURES}
\subsection{Resource availability}
\subsubsection{Lead contact}
For additional information or resource requests, please get in touch with the lead
contact, Xiaolong Yang: yangxl@cqu.edu.cn.

\subsubsection{Materials availability}
This study did not generate new unique materials.

\subsubsection{Data and code availability}
The data generated in this study are available from the lead contact upon reasonable
request.

\subsection{Computational Methods}

All first-principles calculations were implemented in the Vienna Ab initio Simulation Package (VASP) \cite{PhysRevB.48.13115,PhysRevB.54.11169} with the projector augmented wave potential \cite{PhysRevB.50.17953}.  The generalized gradient approximation (GGA) with the Perdew-Burke-Ernzerhof functional (PBE) was used for the exchange-correlation functional \cite{PhysRevLett.100.136406,Le_2012}. The topological surface states were calculated by the Wanniertools package \cite{WU2017}, and the irreducible representations of phonon branches were obtained by the PhononIrep package \cite{zhang2022phonon}. The harmonic interatomic force constants (IFCs) were calculated using 4$\times$4$\times$4 supercells within a finite displacement method, as implemented in the open source software Phonopy package \cite{TOGO20151}. With the same method, the third-order IFCs were obtained using the Thirdorder script \cite{LI20141747}, considering the atomic interactions up to the tenth-nearest neighbors with 4$\times$4$\times$4 supercells. To check the importance of four-phonon scattering, the fourth-order FCs are needed, which were calculated considering the second nearest neighbors with 4$\times$4$\times$4 supercells, by utilizing the Fourthorder code \cite{han2022fourphonon}. To obtain the accurate $\kappa_{L}$, the linearized Boltzmann transport equation (BTE) was exactly solved within an iterative scheme using a 16$\times$16$\times$16 $\bf{q}$-points, by employing a modified version of ShengBTE package \cite{LI20141747,han2022fourphonon}. Here the phonon lifetimes were calculated by considering the isotope and three-phonon scattering. 

The phonon scattering rate due to three-phonon processes can be expressed as 
\begin{equation}
\frac{1}{\tau_{3, \lambda}} =\frac{1}{N}\left( \sum_{\lambda^{\prime}\lambda^{\prime\prime}}^{+}{\Gamma_{\lambda\lambda^{\prime}\lambda^{\prime\prime}}^+}+\sum_{\lambda^{\prime}\lambda^{\prime\prime}}^{-}{\Gamma_{\lambda\lambda^{\prime}\lambda^{\prime\prime}}^-} \right), \tag{Equation 1}
\end{equation}

\begin{equation}
\Gamma_{\lambda\lambda^{\prime}\lambda^{\prime\prime}}^+=\frac{\hbar\pi}{4}\frac{f^{\prime}_0-f^{\prime\prime}_0}{\omega_{\lambda}\omega_{\lambda\prime}\omega_{\lambda^{\prime\prime}}}\left| V_{\lambda\lambda^{\prime}\lambda^{\prime\prime}}^+ \right|^2 
\delta(\omega_{\lambda}+\omega_{\lambda\prime}-\omega_{\lambda^{\prime\prime}}), \tag{Equation 2}
\end{equation}

\begin{equation}
\Gamma_{\lambda\lambda^{\prime}\lambda^{\prime\prime}}^-=\frac{\hbar\pi}{4}\frac{f^{\prime}_0+f^{\prime\prime}_0+1}{\omega_{\lambda}\omega_{\lambda\prime}\omega_{\lambda^{\prime\prime}}}\left| V_{\lambda\lambda^{\prime}\lambda^{\prime\prime}}^- \right|^2 
\delta(\omega_{\lambda}-\omega_{\lambda\prime}-\omega_{\lambda^{\prime\prime}}), \tag{Equation 3}   
\end{equation}
where $N$ is the number of $\textbf{q}$-points in the Brillouin zone, and $\hbar$ is the reduced Planck constant. $f_0(\omega_{\lambda})$ represents the occupation number of phonon with frequency $\omega$. $\Gamma_{\lambda\lambda^{\prime}\lambda^{\prime\prime}}^+$ corresponds to absorption processes involving one phonon with the combined energy of two phonons, and $\Gamma_{\lambda\lambda^{\prime}\lambda^{\prime\prime}}^-$ describes emission processes in which the energy of one incident phonon is split among two phonons.

As an important quantity determining the three-phonon scattering rate, the scattering matrix elements are given by
\begin{equation}
V_{\lambda\lambda^{\prime}\lambda^{\prime\prime}}^{\pm}=\sum_{i\in u.c.}\sum_{j,k}\sum_{\alpha\beta\gamma}\phi^{\alpha\beta\gamma}_{ijk}
\frac{e_{\lambda}^{\alpha}(i)e_{p^{\prime},{\pm}q^{\prime}}^{\beta}(j)e_{p^{\prime\prime},-q^{\prime\prime}}^{\gamma}(k)}{\sqrt{M_i M_j M_k}}, \tag{Equation 4}
\end{equation}
which depend on the normalized eigenfunctions $e_{p,q}$ of the three phonons involved and on the anharmonic IFCs $\phi^{\alpha\beta\gamma}_{ijk}=\frac{\partial^{3}E}{{\partial}r_{i}^{\alpha}{\partial}r_{i}^{\beta}{\partial}r_{i}^{\gamma}}$, with $i$, $j$, and $k$ representing atomic indices. $r_{i}^{\alpha}$ and $M_{i}$ denote the $\alpha$ component of the displacement from the equilibrium position and the mass of the $i$th atom, respectively. $e_{\lambda}^{\alpha}(i)$ is the $\alpha$ component of the eigenfuinction of mode $\lambda$ at the $i$th atom.

The weighted phase space characterizes the three-phonon scattering events that satisfy phonon energy and momentum conservation conditions, which is expressed as
\begin{equation}
 W_{\lambda}^{\pm}=\frac{1}{2N}\sum_{\lambda^{\prime}\lambda^{\prime\prime}} 
 \begin{Bmatrix}
 2\left(f^{\prime}_0-f^{\prime\prime}_0 \right)\\
 f^{\prime}_0+f^{\prime\prime}_0+1
 \end{Bmatrix}
 \frac{\delta \left(\omega_{\lambda}\pm\omega_{\lambda\prime}-\omega_{\lambda^{\prime\prime}} \right)}{\omega_\lambda{\omega_{\lambda\prime}}{\omega_{\lambda^{\prime\prime}}}}. \tag{Equation 5}
\end{equation}
 for absorption (+) and emission (-) processes, respectively. 

\section{Acknowledgments}
X.Y. acknowledges support from the Natural Science Foundation of China (NSFC) (Grant Nos. 12374038 and 12147102.) and the Chongqing Natural Science Foundation (Grant No. CSTB2022NSCQ-MSX0834). X.J. acknowledges support form the Foundation of Chongqing Normal University (Grant No. 23XLB015). D.-S. Ma. acknowledges support from the NSFC (Grants No. 12204074) and the China National Postdoctoral Program for Innovative Talent (Grant No. BX20220367). X.L. acknowledge support from the National Key R$\&$D Program of China (Grant No. 2018YFC1900500), the Foundation Research Fund for NSFC (Grant No. U1902217), and Chongqing Outstanding Youth Project (Grant No. CSTC2019JCYJJQX0024). Simulations have been performed on Hefei advanced computing center.

\section{AUTHOR CONTRIBUTIONS}
 X.Y., X.J, and D.-S.Ma. conceived and designed the study. X.J, X.Y.D., and P.Y performed the ab initio calculations. X.Y. and X.L. supervised the research. X.J and X.Y. coordinated the writing of the manuscript with discussion and input from all authors.

\section{DECLARATION OF INTERESTS}
The authors declare no competing interests.

\end{document}